\documentclass[letterpaper,8pt]{extarticle}  % "extarticle" gives more global font size options. use 8pt with Montserrat and 10pt with EBGaramond
\usepackage{import}
\usepackage[left=.65in,top=.65in,bottom=.65in,right=.65in]{geometry}
\usepackage{lipsum} % to generate text
\usepackage{graphicx}
\usepackage{siunitx}
\usepackage{amsmath}

\title{\huge Spatially Selective Acoustic Pressure Reporting Using Antibubbles}

\author{
  Nicolas Moreno Gomez\textsuperscript{1,2}\footnotemark[1]\footnotemark[2], 
  Athanasios G. Athanassiadis\textsuperscript{1,2}\footnotemark[1]\footnotemark[2],\\
  Fabian Reuter\textsuperscript{3}, Hendrik Reese\textsuperscript{3,4}, 
  Helen M. Jade\textsuperscript{1},\\
  Albert Poortinga\textsuperscript{5}, Claus-Dieter Ohl\textsuperscript{3}, 
  Peer Fischer\textsuperscript{1,2,6,7}\footnotemark[1]
}

\date{}

\begin{document}
\maketitle

% Define footnotes after the title
\footnotetext[1]{*Corresponding authors: nmoreno@uni-heidelberg.de, thanasi@uni-heidelberg.de, peer.fischer@mr.mpg.de.}
\footnotetext[2]{These authors contributed equally to this work.}

% Affiliations
\noindent
\textsuperscript{1}Institute for Molecular Systems Engineering and Advanced Materials, Heidelberg University, Im Neuenheimer Feld 225, D-69120 Heidelberg, Germany.\\
\textsuperscript{2}Max Planck Institute for Medical Research, Jahnstraße 29, D-69120 Heidelberg, Germany.\\
\textsuperscript{3}Department for Soft Matter, Institute for Physics, Otto-von-Guericke University, Magdeburg 39106, Germany.\\
\textsuperscript{4}Lehrstuhl Strömungsmechanik \& Strömungstechnik, Institut für Strömungstechnik und Thermodynamik, Otto-von-Guericke-University, Magdeburg 39106, Germany.\\
\textsuperscript{5}Polymer Technology Group, Eindhoven University of Technology, De Rondom 70, Eindhoven 5612 AZ, The Netherlands.\\
\textsuperscript{6}Center for Nanomedicine, Institute for Basic Science (IBS), Seoul 03722, Republic of Korea.\\
\textsuperscript{7}Department of Nano Biomedical Engineering (NanoBME), Advanced Science Institute, Yonsei University, Seoul 03722, Republic of Korea.\\

%%%%%%%%%%%%%%%%%%%%%%%%%%%%%%%%%%%%%%%%%%%%%%%%%%%%%%%%%%%%%%%%%%%%%
% Abstract
%%%%%%%%%%%%%%%%%%%%%%%%%%%%%%%%%%%%%%%%%%%%%%%%%%%%%%%%%%%%%%%%%%%%%
\section{Abstract}

%\begin{doublespacing}
%\begin{linenumbers}
Ultrasound offers promising applications in biology and chemistry, but quantifying local ultrasound conditions remains challenging due to the lack of non-invasive measurement tools. We introduce antibubbles as novel optical reporters of local ultrasound pressure. These liquid-core, air-shell structures encapsulate fluorescent payloads, releasing them upon exposure to low-intensity ultrasound. We demonstrate their versatility by fabricating antibubbles with hydrophilic and hydrophobic payloads, revealing payload-dependent encapsulation efficiency and release dynamics. Using acoustic holograms, we showcase precise spatial control of payload release, enabling visualization of complex ultrasound fields. High-speed fluorescence imaging reveals a gentle, single-shot release mechanism occurring within 20-50 ultrasound cycles. It is thus possible to determine via an optical fluorescence marker what the applied ultrasound pressure was. This work thereby introduces a non-invasive method for mapping ultrasound fields in complex environments, potentially accelerating research in ultrasound-based therapies and processes. The long-term stability and versatility of these antibubble reporters suggest broad applicability in studying and optimizing ultrasound effects across various biological and chemical systems.
\noindent

%\begin{multicols*}{2} % if you want two columns
%%%%%%%%%%%%%%%%%%%%%%%%%%%%%%%%%%%%%%%%%%%%%%%%%%%%%%%%%%%%
\section{Introduction}
%%%%%%%		 Main Text			%%%%%%% 

Numerous emerging techniques in biology and chemistry harness ultrasound-induced thermal and mechanical effects, offering deep penetration, high biocompatibility, and tight spatiotemporal control.\cite{athanassiadis:2022} Recent studies indicate that even low-intensity ultrasound can produce useful effects, minimizing collateral damage in biological systems.\cite{du2022impact,jiang2018review} However, understanding the fundamental mechanisms of these effects, particularly in small systems, remains challenging due to difficulties in quantifying local ultrasound conditions.\cite{leskinen:2012, fontana:2021}

Unlike light, where fluorophores can act as passive intensity reporters,\cite{mandal2023series,zheng2014ultra} no such non-invasive tools exist for low-intensity ultrasound. Current measurement techniques are typically invasive or cannot be easily integrated into many experimental systems. This limitation hinders the optimization of existing applications and the development of new ones in fields ranging from targeted therapies to nanoscale material manipulation.

To address this challenge we show how antibubbles can serve as nearly-passive optical reporters of local ultrasound pressure. Antibubbles—liquid droplets surrounded by a stabilized air shell\cite{zia2022advances}—can carry and release liquid payloads upon ultrasound excitation.\cite{moreno2023antibubbles,kotopoulis2022formulation} Compared to conventional surface-loaded microbubbles, antibubbles offer larger payload capacities and respond to lower acoustic pressures, making them ideal for studying low-intensity ultrasound effects.

In this work, we demonstrate the use of acoustic holograms to precisely control the spatial release of payloads from antibubbles, enabling visualization of ultrasound effects across various length and time scales. Additionally, we showcase the ability to observe payload release dynamics at the microscale using high-speed imaging, similar to traditional methods that focus on visualizing effects during ultrasound exposure. Our approach offers a rapid, one-shot method to control payload release in space and time, representing a significant step towards understanding and quantifying the chemical and biological effects induced by ultrasound. We propose that antibubbles, which can be fabricated to respond to a particular pressure, can readily be embedded in soft materials and biological systems to serve as in-situ reporters for ultrasound applications.

%%%%%%%%%%%%%%%%%%%%%%%%%%%%%%%%%%%%%%%%%%%%%%%%%%%%%%%%%%%%
\section{Results and Discussion}
%%%%%%%%%%%%%%%%%%%%%%%%%%%%%%%%%%%%%%%%%%%%%%%%%%%%%%%%%%%%
%%%%%%%%%%%%%%%%%%%%%%%%%%%%%%%%%%%%%%%
\subsection*{Payload-dependent antibubble behavior revealed through fabrication and characterization}
%%%%%%%%%%%%%%%%%%%%%%%%%%%%%%%%%%%%%%%%

\begin{figure*}
\begin{center}
\includegraphics[width=17.4cm]{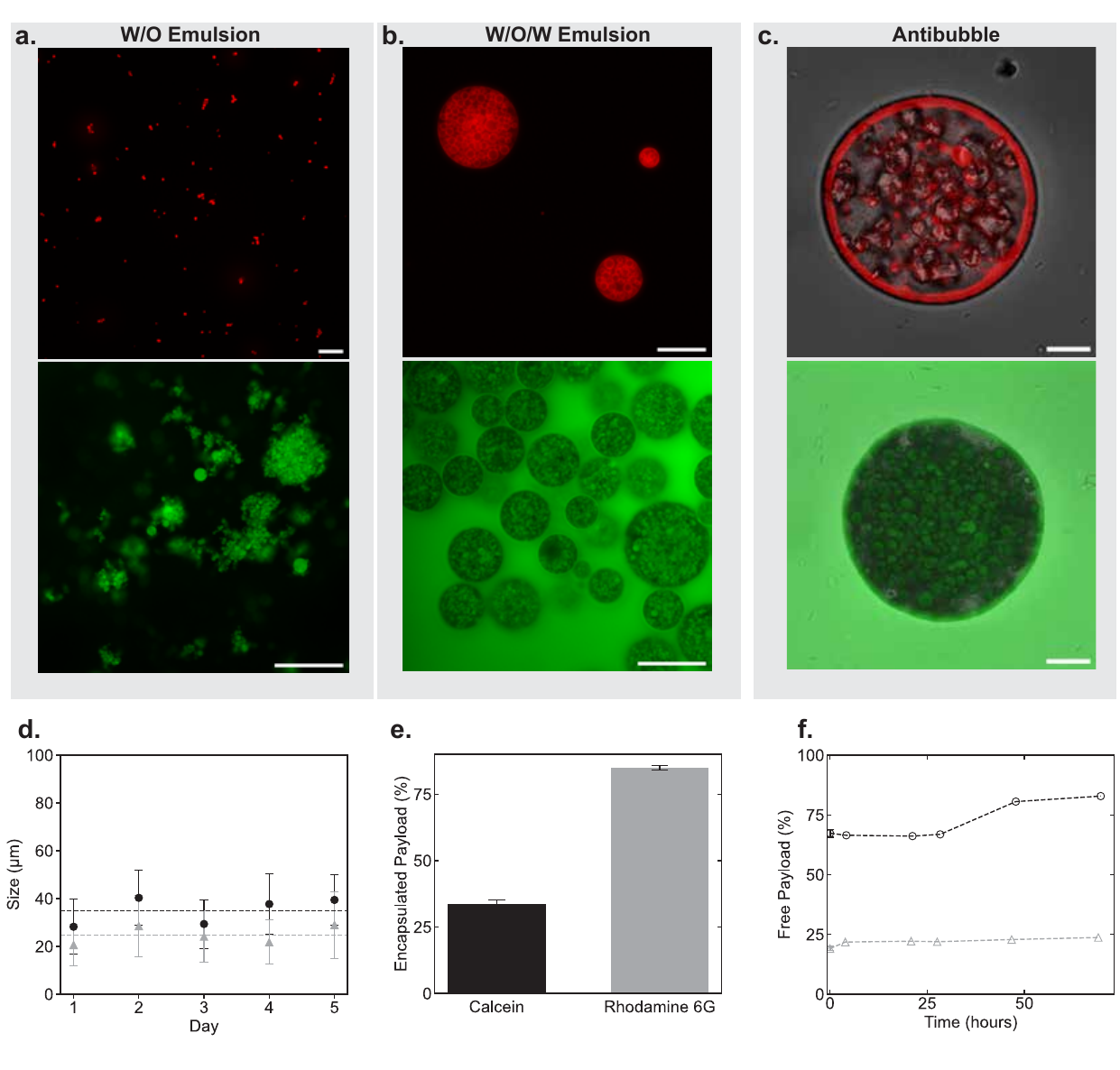}
\caption{Payload-specific encapsulation of antibubbles with similar shell composition but different payloads. (a,b) Fluorescence microscopy of W/O and W/O/W emulsion droplets respectively (scale bars: 50 µm). Top images correspond to Rhodamine 6G and bottom to Calcein loaded formulation. (c) Confocal cross-section of an antibubble (scale bar: 10 µm). (d-f) Despite identical composition, Calcein (hydrophilic) and Rhodamine 6G (hydrophobic) loaded antibubbles differ in: (d) their size distributions over time, (e) encapsulation efficiency, and (f) the cumulative release, demonstrating higher leakage for Calcein and superior encapsulation for Rhodamine 6G.}
%Should  y-axis of subfigure d rather state "mean radius" or "mean diameter" 
\label{fig1ABS}
\end{center}
\end{figure*}

We employed an established method to create two types of antibubbles, \cite{moreno2023antibubbles, kotopoulis2022formulation}each encapsulating a different payload. The fabrication involves first the integration of the payload via a water-in-oil (W/O) Pickering emulsion, followed by templating the structure using a W/O/W emulsion, and finalized by forming the antibubble after freeze-drying and reconstituting in an aqueous solution.\cite{moreno2023antibubbles} Fluorescence microscopy images reveal the successful formation of W/O and W/O/W emulsions (Figure~\ref{fig1ABS}, panels \textbf{a.} and \textbf{b.}). The W/O emulsion, stabilized using only hydrophobic fumed silica (Evonik, R972), encapsulates Rhodamine 6G at a concentration of \SI{1.06}{\milli\mole\per\liter} as well as Calcein at a concentration \SI{120}{\milli\mole\per\liter}. This difference in concentrations shows the loading capability of antibubbles that is limited by the solubility of the payload in water. The W/O/W is stabilized using a 2:1 mixture of hydrophilic (Evonik, Aerosil 200) to hydrophobic (Evonik, R972) silica for the external shell.\cite{EvonikAerosil} The W/O/W emulsion images show the hydrophilic Calcein dye is also found in the aqueous phase, whereas Rhodamine 6G is only seen in the core of the antibubbles. Upon freeze-drying and reconstitution in aqueous sodium chloride, the antibubbles are transferred and resuspended in a xanthan gum solution (\SI{0.5}{\percent} wt). Confocal microscopy confirms the antibubble formation but reveals external fluorescence for the Calcein-loaded antibubbles, even after the fabrication and cleaning steps. 

Size measurements over five days showed consistent mean sizes for both Calcein and Rhodamine 6G-loaded antibubbles (Figure~\ref{fig1ABS}, Panel \textbf{d.}), with Calcein exhibiting a higher mean value, but both demonstrating sufficient stability to rule out spontaneous destruction. Remarkably, the Calcein sample, fabricated three years prior to this measurement, yielded the same size as that measured immediately after fabrication,\cite{moreno2023antibubbles} indicating exceptional long-term shelf stability of the freeze-dried antibubble powder. While both types of antibubbles showed the same overall long-term stability, the encapsulation efficiency and the dye diffusion through the shell differed (see Figure~\ref{fig1ABS}, \textbf{e.} and \textbf{f.}). Rhodamine 6G antibubbles showed high encapsulation (\SI{>80}{\percent}) and low leakage (\SI{20}{\percent} free dye), while Calcein exhibited low encapsulation (\SI{35}{\percent}) and high leakage (\SIrange{70}{80}{\percent} free dye). This payload-dependent behavior reveals that payload-particle interactions significantly influence antibubble performance, beyond previously reported factors of size and shell composition. The tendency of the hydrophilic Calcein to leak suits rapid, controlled release applications, while the superior encapsulation of the hydrophobic Rhodamine 6G allows sustained delivery. These findings guide our subsequent experiments, in which we show controlled release of Calcein-loaded antibubbles using acoustic holograms and low-intensity ultrasound, and Rhodamine 6G release using high-speed fluorescence imaging.

 \subsection*{Spatially-specific pressure reporting in the field of an acoustic hologram}

\begin{figure*}
\begin{center}
\includegraphics[width=17.4cm]{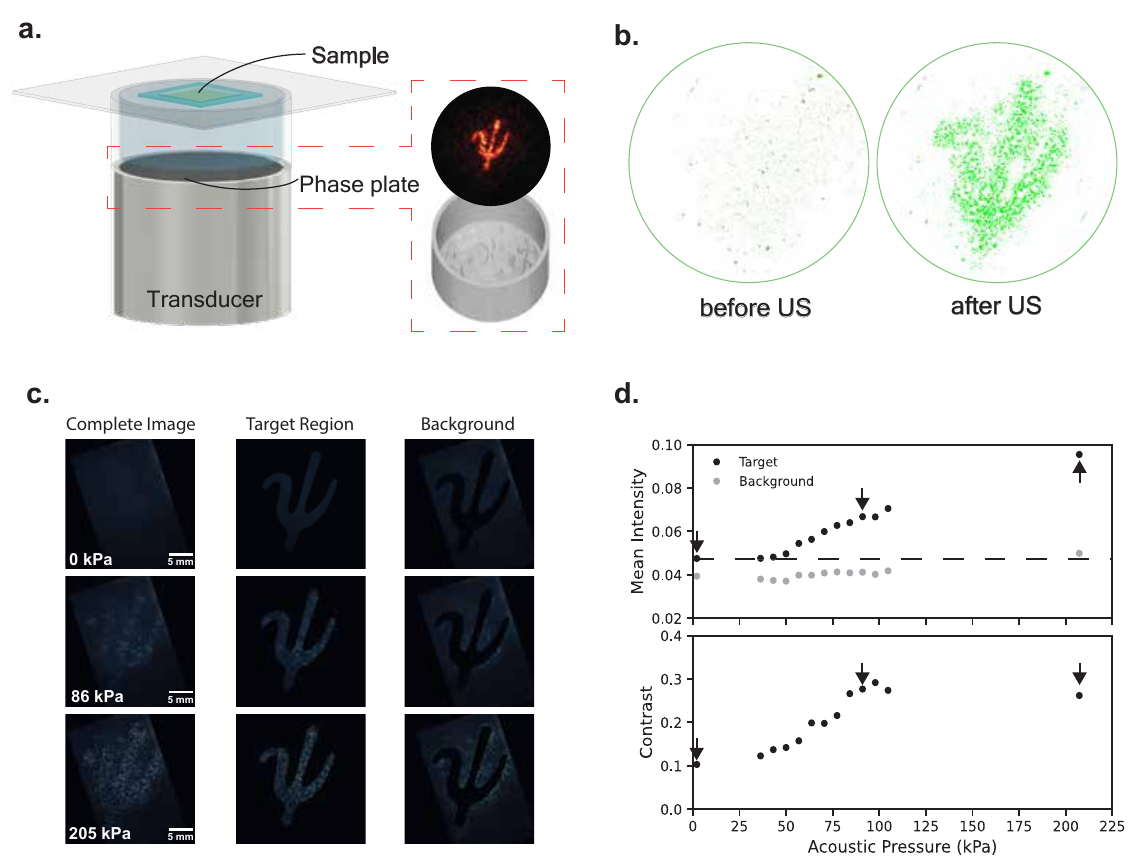}
\caption{Spatially controlled and pressure-dependent payload release from different Calcein-loaded antibubble formulations using acoustic holograms. (a) Schematic of the experimental setup. (b) Processed fluorescence images before and after the ultrasound pulse, demonstrating spatially controlled, one-shot release from antibubbles with hydrophilic external shells. (c,d) Pressure-dependent, stepwise release from antibubbles with hydrophobic external shells: (c) Processed images (grayscale and pseudo-color) of the target area and background intensity at three different pressures. (d) Mean intensity and contrast plots across 13 pressure levels, with arrows indicating the pressures shown in (c). As the incident pressure increases, antibubble release clearly delineates the hologram focal pattern, until the background pressure becomes too high, causing additional release outside of the central focal region. This is reflected in a decrease in contrast above 100 kPa.}
\label{fig2holo}
\end{center}
\end{figure*}

To evaluate the spatial specificity of the pressure reporting by antibubbles, we exposed Calcein-loaded antibubbles to a \SI{1}{\mega\Hz} ultrasound field that was spatially patterned in the shape of the Greek letter $\psi$ using an acoustic hologram (Figure~\ref{fig2holo}).\cite{melde:2016} The hologram was 3D printed from a UV-cured polymer (Vero Clear, Stratasys) onto the base of a water-filled cylindrical container that could be mounted on top of an ultrasound transducer with center frequency \SI{1.0}{\mega\hertz} (Precision Acoustics). The pressure field was verified with a 2D hydrophone scan (Figure~\ref{fig2holo}). The container was filled with water and an evenly-distributed antibubble sample of \SI{90}{\micro\liter} was filled into a \SI{125}{\micro\liter} chamber (Gene Frame; Thermo Scientific) mounted on a glass substrate. The chamber was sealed with an acoustically-transparent polyester film, making sure to not trap air bubbles, and then the sample was inverted and placed on the water-filled hologram container, so that sound could propagate up from the hologram, through the water, and into the antibubble sample (Figure~\ref{fig2holo} \textbf{a.}). % Acoustic reflections off the glass substrate at the top lead to constructive interference that double the pressure at the sample compared to the free-field pressure.

For accurate spatial reporting, the antibubbles needed to be stabilized against motion that can arise from gravity (buoyancy) or acoustic effects including streaming and radiation forces. We stabilized the antibubbles in two ways. First, the Calcein-loaded antibubbles were suspended in an aqueous solution containing 0.4 wt\% xanthan gum, which increases the viscosity of the suspension, stabilizing the bubbles from general movements. Second, the samples were exposed to short bursts (100 cycles) of ultrasound. This length was sufficient for the maximum pressure to be reached in the hologram output field, without inducing significant streaming or long-range bubble motion due to radiation forces.

When the antibubble sample was exposed to the patterned ultrasound field, those antibubbles in high-pressure regions released their payloads while those in low-pressure regions did not, demonstrating the spatial selectivity of the release process in our gel. We visualized the release using fluorescence imaging of the Calcein before and after a \SI{205}{\kilo\Pa} peak-pressure, 100-cycle ultrasound pulse at \SI{1}{\mega\hertz} (Figure~\ref{fig2holo} \textbf{b.}). Before the ultrasound, the overall fluorescence levels are very low, reflecting the quenched Calcein in the antibubble cores. After the ultrasound, on the other hand, the $\psi$ pattern is clearly visible, revealing regions where the ultrasound pressure exceeded the release threshold of the antibubbles. 

Because payload release from this antibubble formulation occurs in a single shot, this antibubble gel reports pressures that exceed the release threshold for the formulation. However, formulations with a hydrophobic shell are capable of cumulative release over a broader range of acoustic pressures.\cite{moreno2023antibubbles} Therefore, to further explore pressure-dependent release dynamics, we fabricated a formulation with a different mean size and hydrophobic shell composition, capable of incremental release. We demonstrated this by gradually increasing the applied pressure from under \SI{1}{\kilo\Pa} to \SI{200}{\kilo\Pa}, and quantifying the fluorescence in the reporter gel (Figure~\ref{fig2holo} \textbf{c.} and \textbf{d.}). At pressures below \SI{50}{\kilo\Pa}, no increase in fluorescence is observed, indicating no release. Above \SI{50}{\kilo\Pa} the fluorescence intensity begins to increase steadily, reflecting the pressure threshold of this formulation around \SI{50}{\kilo\Pa}. At significantly higher pressures of \SI{200}{\kilo\Pa}, more Calcein release was observed but the clarity of the release pattern has deteriorated, suggesting that the background pressure level in the hologram led to a loss of spatial specificity of the antibubble indicator.

Next, we quantified the spatial specificity of the antibubbles indicator by comparing the measured fluorescence signal to the designed pressure field. Figure \ref{fig2holo} \textbf{c.} and \textbf{d.} illustrates the pressure-dependent release with processed images at 0, 86, and \SI{205}{\kilo\Pa}, showing the full image, target area, and background intensity. These images demonstrate a gradual increase in fluorescence with pressure, not only in the target area but also in the background, albeit to a lesser extent. To concretely describe this trend, we calculate the mean image intensity within the target region and background for images acquired across the tested pressure range. The mean target intensity reflects the sensitivity of the indicator to the focal pressure, while to capture the spatial sensitivity we use the contrast between the target and background (see Methods). This analysis revealed a rapid increase in fluorescence intensity in the target area, with a relatively constant background intensity, confirming the spatial selectivity. Only at high pressures, where the background pressure noise is high enough to release the anitbubble payload, did the background intensity begin to increase, causing a slight decrease in contrast. In our system, the contrast increases with driving pressure up to a maximum at \SI{98}{\kilo\Pa}. These ranges however will vary with the signal-to-noise ratio (SNR) of the pressure field. Fluorescent reporters in a high-SNR field can continue to increase contrast to higher pressures, until the background pressure noise is comparable to the release threshold of the antibubbles. This experiment reveals the high sensitivity and contrast of antibubble-based reporters in measuring spatially complex pressure fields generated by acoustic holograms.

\subsection*{Gentle release process revealed by ultra-high-speed imaging}

\begin{figure*}
\begin{center}
\includegraphics[width=17.4cm]{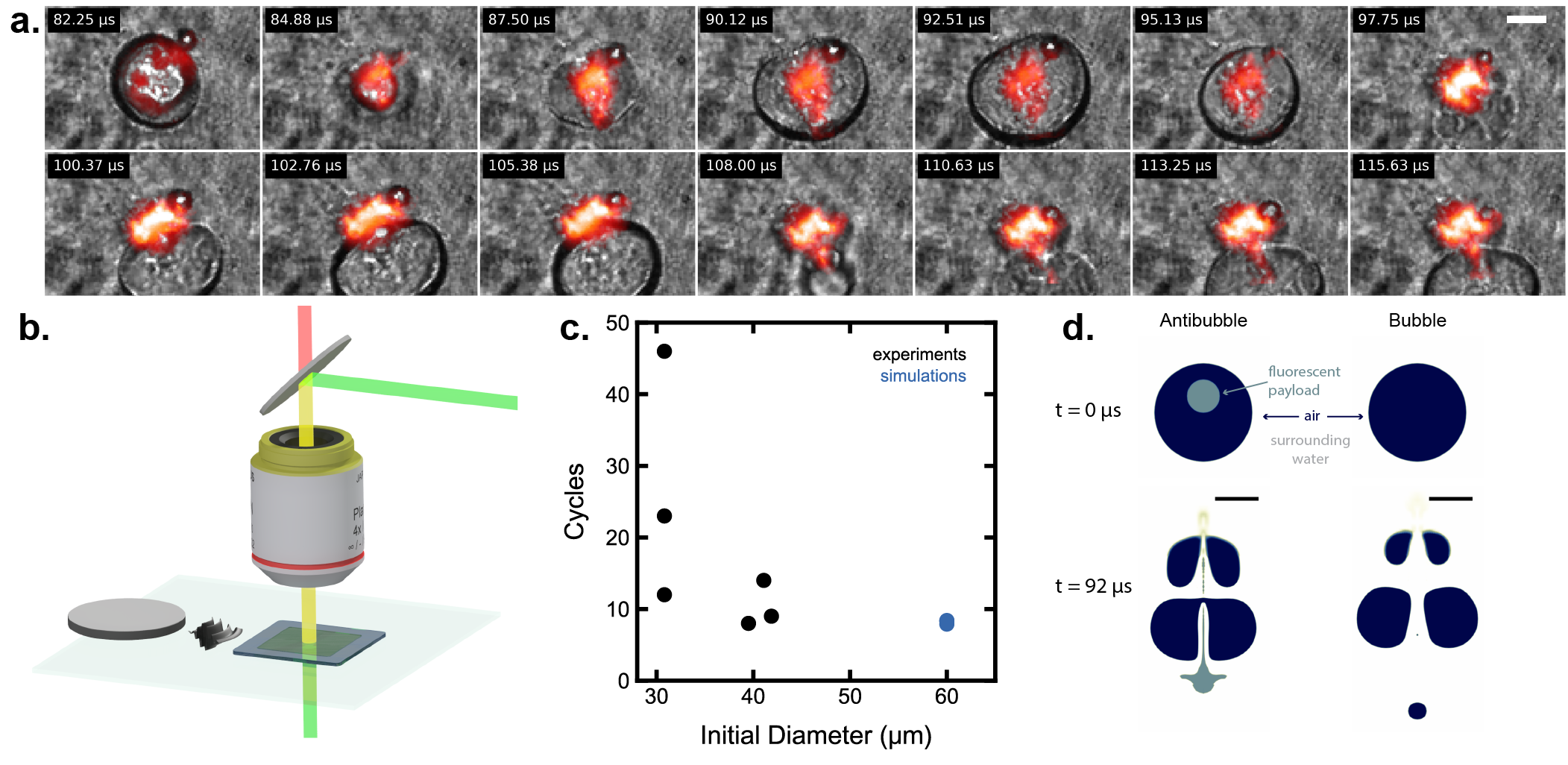}
\caption{High-speed imaging and simulation reveal gentle, rapid payload release from Rhodamine 6G-loaded antibubbles. (a) Time-lapse overlay of brightfield and fluorescence images showing antibubble oscillation and single-bolus payload release.(b) Schematic of the custom high-speed microscopy setup for simultaneous brightfield and fluorescence imaging. (c) Plot of release timing vs. initial antibubble diameter, showing release within 50 cycles. (d) Comparison of simulated antibubble (left) and regular bubble (right) dynamics, demonstrating the unique ejection behavior of antibubbles in agreement with experiments. Scalebars in (a) and (d) are \SI{30}{\micro\meter}.}
\label{fig:hispeed}
\end{center}
\end{figure*}

To validate that antibubbles serve as a passive reporter, we observed the release dynamics from Rhodamine 6G-loaded antibubbles using high-speed microscopy, simultaneously observing the bubble dynamics and the release of the fluorescent payload from antibubbles, as shown in Figure~\ref{fig:hispeed} \textbf{a.}. We built a custom high-speed microscope using two synchronized and co-registered high-speed cameras. The antibubble sample was loaded onto a glass substrate in the same manner as described above, and the sample was excited at \SI{90}{\kilo\hertz} via guided waves in the substrate (Figure~\ref{fig:hispeed} \textbf{b.}). The sample was illuminated from below with nanosecond pulses from a \SI{532}{\nano\meter} laser, and was imaged with a 20$\times$ objective (Olympus UPlan S Apo 0.75). The transmitted and fluorescence light collected by the objective was split by a dichroic mirror, with the fluorescence signal transmitted to one high-speed camera and the brightfield signal reflected to the second high-speed camera (Figure~\ref{fig:hispeed} \textbf{b.}). This setup allowed us to observe both payload release (fluorescence) and the associated bubble dynamics (brightfield).

High-speed imaging revealed that our antibubble reporters release their payloads at once (Figure~\ref{fig:hispeed} \textbf{a.}), consistent with our use of a single-shot formulation. We were able to track the exact release timing for six different antibubbles and found that the release always occurred within 50 cycles after the onset of ultrasound (Figure~\ref{fig:hispeed} \textbf{c.}). In most cases, the payload was released in fewer than 20 cycles of ultrasound, with a tendency of larger antibubbles to release their payloads faster. This likely arises because the resonance frequency of the larger antibubbles here was closer to the ultrasound frequency used, leading to stronger oscillations (according to Rayleigh's collapse time \cite{}, the resonant bubble size diameter is about \SI{40}{\micro\meter}). The rapid, single-shot release of the payload from the antibubles is advantageous for passive indication, because it indicates that a low number of cycles is needed to evaluate the pressure field, consistent with our 100-cycle results on the hologram. With a single shot of so few cycles, very little acoustic energy is introduced to the system, limiting the other acoustic effects (e.g. as heating, streaming, and radiation forces) that could disturb a sample of interest.

In addition to the release timing, our imaging setup revealed the bubble dynamics associated with the release process (Figure~\ref{fig:hispeed} \textbf{a.}), further supporting the fact that release occurs with limited impact to the surrounding medium. Much like for bubbles, the gas volume in the antibubbles oscillated between expansion and contraction in the presence of ultrasound, growing in amplitude with each cycle. The antibubble oscillations quickly become nonlinear and the bubble quickly begins to collapse. When the bubble collapses to a region smaller than the droplets comprising the core, the gas phase separates from the core upon re-expansion, ejecting the core as a single bolus, and continuing its oscillations independently. In many cases, the bubble continued to migrate away from the ejected payload, likely because of radiation forces acting on the bubble.

Surprisingly, we did not observe any evidence of jetting-based release or significant bubble fragmentation, which is typical of release from surface-loaded microbubbles. The release behavior of antibubbles is in fact much gentler than that from surface-loaded bubbles. Because surface-loaded bubbles require payloads to be conjugated to their stabilizing shells, release can typically only occur via fragmentation of the bubbles and shells, which requires much higher pressures and is a much more violent process. By contrast, the migration of the gas phase away from the core occurs in the course of stable cavitation.

One important factor that may be strongly influencing the release dynamics from the antibubbles is the stabilizing fumed silica shell. In some cases, we observed the payload adhering to the outer walls of the antibubble, likely because of the hydrophobicity of the fumed silica. These interactions were strong enough that in some cases the droplets even remained stuck to the bubble walls for multiple cycles after release, until hydrodynamic forces helped to release them from the bubble, though this effect may also be caused by surface tension. While we only observed such behavior in a minority of measurements, it nonetheless indicates that the fumed silica can mediate or potentially even inhibit the release of the payload in certain conditions. Moreover, it hints that the payload may retain its stabilizing shell even after release. While we did not qualitatively observe any long-time inhibition of diffusion because of this shell, the fumed silica may play an additional role in mediating the diffusion of payloads after release - a behavior that could be helpful or disruptive depending on the application. These results therefore highlight the importance of the fumed silica formulation on the release dynamics - a relationship that requires further investigation. 

Finally, to test our hypothesis that the bubble dynamics with a liquid core lead to the sudden ejection behavior we have observed with the antibubbles, we modeled the payload release from the antibubbles using OpenFOAM, a numerical framework for computational fluid dynamics.\cite{OpenFOAM-Foundation2021} The simulation consisted of a spherical water droplet inside of a spherical air bubble surrounded by water. The simulations were axisymmetric, which eased the computational cost by computing the dynamics in a 2D slice through the center of the antibubble that, when rotated, represents the full 3D structure. For comparison, we also simulated a gas bubble without a liquid core. Axisymmetry is an appropriate assumption for spherical bubble dynamics, and as we describe below, produces physically meaningful results in good agreement with our experimental observations. In both simulations, the antibubbles were exposed to a spatially uniform, time-varying pressure oscillating at \SI{100}{\kilo\hertz} and an amplitude \SI{10}{\kilo\Pa} from atmospheric pressure. In line with our experiments, the simulations predicted a rapid release process, with the payload released as a single droplet within 10 cycles. Because the simulated bubbles were larger than those in the experiments, they were closer to resonance and therefore released their payloads sooner, continuing the experimental trend we observed of faster release from larger bubbles (Figure~\ref{fig:hispeed} \textbf{c.}). Additional agreement can be observed between the simulations and experimental observations. For instance, the gas phase migrates in the opposite direction of the liquid phase before and after the ejection, as observed in our experiments. Moreover, the payload is ejected within a single collapse cycle of the bubble and emerges as a single droplet. The subsequent fluid dynamics surrounding the bubble lead to a thin trailing jet of the payload pointing into the bubble, and a disk-shaped wing spreading perpendicular from the axis of symmetry. The same features can be observed in the experimental recordings in the last frame of Figure~\ref{fig:hispeed} \textbf{a.}. 

Several similarities and differences can be observed between the bubble and antibubble dynamics that highlight the possibility of gentler release from antibubbles. In particular, the general bubble breakup processes are similar after 10 cycles, with the gas bubbles ultimately fragmenting. However, the fact that the liquid core of the antibubble is released by this point is in stark contrast to what would happen with surface-loaded microbubble-based delivery. In the case of surface-loaded bubbles, the high stability of the shell leads to higher pressures required for disruption and bubble fragmentation,\cite{stride2019nucleation,shakya2024ultrasound} which can be associated with additional detrimental effects.\cite{wang2013ultrasound,delaney2022making} Moreover, the release of a surface-conjugated payload would require exposure to more intense or longer pulses in order to release the conjugated payload.\cite{shakya2024ultrasound} Finally, in the case of co-injected payloads, the bubbles would need to undergo stronger oscillations to induce sonoporation via mechanical effects or strong fluid jets\cite{stride2019nucleation,shakya2024ultrasound} that can also be associated with unwanted side-effects.

Some differences can be observed between the simulations and experiments, particularly that the bubble oscillations in experiments grow more quickly, and that the payload sometimes attaches to the outer surface of the bubble for multiple cycles after release from the bubble. Such behavior is likely due to the adhesion of the silica stabilizing layer, which we did not model in the simulations. Nonetheless, despite the simplifications involved in the simulations, the strong agreement between the simulations and experiments highlights that the process is dominated by the liquid-gas dynamics, and that such simulations can be used as a tool for better understanding and optimization of the release process.
%%%%%%%%%%%%%%%%%%%%%%%%%%%%%%%%%%%%%%%%%%%%%%%%%%%%%%%%%%%%
\section{Conclusion}
%%%%%%%%%%%%%%%%%%%%%%%%%%%%%%%%%%%%%%%%%%%%%%%%%%%%%%%%%%%%
In summary, our simulations and dual fluorescence-brightfield high-speed microscopy provide unique insight into the coupling of the release process and the bubble dynamics. These techniques allow us to identify a gentler payload release process than would be possible with conventional surface-loaded microbubbles. Most notably, antibubbles tended to release their payloads as a single bolus during a single compression-expansion cycle of the gas. This typically occurred within 20 cycles, highlighting the limited energy needed to evaluate the spatial structure of a field. These characteristics make the antibubbles a particularly effective and low-impact fluorescent probe of a pressure field in confined and complex geometries. Due to their small size and exceptional stability the antibubbles presented herein can serve as an in-situ reporter of pressure fields. We believe that the concept of an optical intensity marker of US pressure -- realized here with dye-loaded antibubbles -- will be especially useful for in-vitro\cite{melde2024ultrasound} and in-vivo applications of ultrasound.

%%%%%%%%%%%%%%%%%%%%%%%%%%%%%%%%%%%%%%%%%%%%%%%%%%%%%%%%%%%%
\section{Materials and Methods}
%%%%%%%%%%%%%%%%%%%%%%%%%%%%%%%%%%%%%%%%%%%%%%%%%%%%%%%%%%%%

\subsection{Antibubble Fabrication and Characterization}

The general fabrication of the different antibubbles was performed as described elsewhere \cite{kotopoulis2022formulation}. Calcein antibubbles were fabricated following the conditions described in Ref. \cite{moreno2023antibubbles}. Rhodamine 6G-loaded antibubbles were fabricated as follows: First, the inner aqueous solution was prepared by dissolving in \SI{16.7}{\g} type I water (Millipore), \SI{1.85}{g} maltodextrin and \SI{9.4}{\milli\g} of Rhodamine 6G. Second, \SI{1.041}{\g} of fumed silica Aerosil R972 (Evonik) was dispersed in \SI{33.96}{\g} of cyclohexane using an ultrasonic sonotrode (Hirscher, UP100). Third, the external aqueous phase consisted of \SI{10}{g} of water, \SI{2}{g} of maltodextrin, and \SI{0.021}{\g} of xanthan gum. Then this solution was used to disperse a mixture of Aerosil 200 (\SI{0.1}{\g}) and Aerosil R972 (\SI{0.05}{\g}) in a 2:1 mass ratio using rotor-stator homogenization (Ultraturrax T18, IKA) and an ultrasonic sonotrode (Hirscher, UP100). Next, \SI{8
6}{\g} of the inner aqueous solution was dispersed in \SI{11.4}{\g} of the medium solution using an ultrasonic sonotrode (Hirscher, UP100) for 1\,min, 100\% amplitude and Cycle 1 to form a W/O emulsion. Next, \SI{4}{\g} of the W/O emulsion was combined with \SI{16}{\g} of external solution containing fumed silica by rotor-stator homogenization (Ultraturrax T18, IKA) at 14\,000\,rpm to obtain W/O/W emulsions. The resulting emulsion was then centrifuged at 750\,rcf for 3 min and the cream was scooped out and transferred to a small glass vial. The sample was frozen using liquid Nitrogen and dried for 24\,h at -113.5\,$^{\circ}$C and \SI{0.1}{\milli\bar} (Martin Christ, Alpha 3-4 LSCplus). The resulting dry powder was stored and used after reconstitution in aqueous NaCl 0.3\,wt$\%$. \hfill \break

Aerosil R972 and Aerosil 200 were generously provided by Evonik. Rodamine 6Gwas obtained from Carl Roth. Maltodextrin, cyclohexane, and sodium chloride were purchased from Sigma-Aldrich. All reagents were used without further purification.

\subsection{Microscopy characterization}
Fluorescence microscopy images were obtained using an inverted optical microscope (Zeiss Observer D1) equipped with a fluorescence lamp (Colibri 7) using different magnifications ($2.5\times$, $10\times$, and $40\times$). Confocal microscopy was performed on a Leica XXX (inverted) laser-scanning confocal microscope. All images were processed using the software ImageJ.\hfill \break

Antibubble size distributions were measured using an inverted optical microscope (Zeiss Observer D1) with a $10\times$ objective. Size distributions are based on microscopy images, where ImageJ was used to measure the diameter of antibubbles. The scale was determined using a calibration slide. Then mean size and standard deviation were used to calculate the Polydispersity Index (PDI) as follows:

\begin{equation}
PDI = \left( \dfrac{\text{Standard Deviation}}{\text{Mean Diameter}} \right)^2
\end{equation}\hfill \break

\subsection{Encapsulation Efficiency and Cumulative Release}

For the determination of the Encapsulation Efficiency (EE), a batch of each antibubble formulation was prepared. Specifically, \SI{0
1}{\g} of antibubble powder was mixed with \SI{12.0}{\g} of NaCl solution (0.3\,wt\% for Rhodamine 6G, 2\,wt\% for Calcein). The mixture was then rehydrated for 5 minutes. Following rehydration, The liquid infranatant was removed using a syringe with a needle, and \SI{23}{\milli\L} of sodium chloride with 0.5\,wt\% xanthan gum was added. Next, the rehydrated antibubbles were split into six 2\,mL tubes. The samples were centrifuged at 10,000\,rcf for 5 minutes. Absorbance was measured for three of the six tubes by transferring the infranatant into a cuvette with a syringe, using a UV-Vis-NIR spectrophotometer (Cary 5000, Agilent Technologies Inc.). For the remaining three samples, the infranatant was removed, and 2\,mL of 10\,\% Tween 20 (Carl Roth) were added. These samples were shaken and centrifuged at 15\,000\,rcf for 3.5 minutes to ensure complete release of the antibubble content. Lastly, the infranatant from each sample was transferred into a cuvette using a syringe and measured for absorbance. After obtaining the absorbance values, the EE was calculated to determine the amount of dye encapsulated, using the following formula:

\begin{equation}
\text{EE (\%)} = \left(\frac{A_{\text{max}} - A_\text{s}}~,{A_{\text{max}}} \right)*100
\end{equation}

where $A_{s}$ is the absorbance of the payload in the infranatant after centrifugation and $A_{max}$ is the maximum absorbance possible after destabilizing the structure Tween 20.

The remaining samples were used to assess release stability. For this, absorbance measurements were conducted over 4 days using a UV/Vis spectrophotometer (Cary 5000, Agilent Technologies Inc.). At each time point, a 2\,mL was taken from the sample batch and centrifuged at 10,000\,rcf for 5 minutes. The resulting infranatant was transferred into a Quartz cuvette for the absorbance measurement. The cumulative release (CR) was calculated using the following equation:

\begin{equation}
CR (\%) = \frac{A_\text{s}}{A_\text{max}} *100
\end{equation}

%%%%%%%%%%%%%%%%%%%%%%%%%%%%%%%%%%%%%%%%%%%%%%%%%%%%%%%%%%%%%%%%%%%%%
\section{Acknowledgments}
%%%%%%%%%%%%%%%%%%%%%%%%%%%%%%%%%%%%%%%%%%%%%%%%%%%%%%%%%%%%%%%%%%%%%
This work was supported by the European Research Council under the ERC Advanced Grant Agreement HOLOMAN (no. 788296), Deutsche Forschungsgemeinschaft (DFG, German Research Foundation) under contract OH 75/4-1.

%%%%%%%%%%%%%%%%%%%%%%%%%%%%%%%%%%%%%%%%%%%%%%%%%%%%%%%%%%%%%%%%%%%%%
\renewcommand\refname{References}
%%%%%%%%%%%%%%%%%%%%%%%%%%%%%%%%%%%%%%%%%%%%%%%%%%%%%%%%%%%%%%%%%%%%%
\begin{footnotesize}
\bibliographystyle{unsrt.bst} % abbrvnat or unsrt
\textnormal{\bibliography{References.bib}}

\begin{thebibliography}{10}

\bibitem{athanassiadis:2022}
Athanasios~G. Athanassiadis, Zhichao Ma, Nicolas {Moreno-Gomez}, Kai Melde, Eunjin Choi, Rahul Goyal, and Peer Fischer.
\newblock Ultrasound-{{Responsive Systems}} as {{Components}} for {{Smart Materials}}.
\newblock {\em Chemical Reviews}, 122(5):5165--5208, March 2022.

\bibitem{du2022impact}
Meng Du, Yue Li, Qing Zhang, Jiaming Zhang, Shuming Ouyang, and Zhiyi Chen.
\newblock The impact of low intensity ultrasound on cells: Underlying mechanisms and current status.
\newblock {\em Progress in Biophysics and Molecular Biology}, 174:41--49, 2022.

\bibitem{jiang2018review}
Xiaoxue Jiang, Oleksandra Savchenko, Yufeng Li, Shiang Qi, Tianlin Yang, Wei Zhang, and Jie Chen.
\newblock A review of low-intensity pulsed ultrasound for therapeutic applications.
\newblock {\em IEEE Transactions on Biomedical Engineering}, 66(10):2704--2718, 2018.

\bibitem{leskinen:2012}
Jarkko~J. Leskinen and Kullervo Hynynen.
\newblock Study of {{Factors Affecting}} the {{Magnitude}} and {{Nature}} of {{Ultrasound Exposure}} with {{{\emph{In~Vitro}}}} {{Set-Ups}}.
\newblock {\em Ultrasound in Medicine \& Biology}, 38(5):777--794, May 2012.

\bibitem{fontana:2021}
F.~Fontana, F.~Iberite, A.~Cafarelli, A.~Aliperta, G.~Baldi, E.~Gabusi, P.~Dolzani, S.~Cristino, G.~Lisignoli, T.~Pratellesi, E.~Dumont, and L.~Ricotti.
\newblock Development and validation of low-intensity pulsed ultrasound systems for highly controlled in vitro cell stimulation.
\newblock {\em Ultrasonics}, 116:106495, September 2021.

\bibitem{mandal2023series}
Mrinal Mandal, Hessam~Sepasi Tehrani, Qianhua Mai, Emma Simon, Marie-Aude Plamont, Christine Rampon, Sophie Vriz, Isabelle Aujard, Thomas Le~Saux, and Ludovic Jullien.
\newblock A series of caged fluorophores for calibrating light intensity.
\newblock {\em Chemical Science}, 14(47):13799--13811, 2023.

\bibitem{zheng2014ultra}
Qinsi Zheng, Manuel~F Juette, Steffen Jockusch, Michael~R Wasserman, Zhou Zhou, Roger~B Altman, and Scott~C Blanchard.
\newblock Ultra-stable organic fluorophores for single-molecule research.
\newblock {\em Chemical Society Reviews}, 43(4):1044--1056, 2014.

\bibitem{zia2022advances}
Rabia Zia, Akmal Nazir, Albert~T Poortinga, and Cornelus~F van Nostrum.
\newblock Advances in antibubble formation and potential applications.
\newblock {\em Advances in Colloid and Interface Science}, 305:102688, 2022.

\bibitem{moreno2023antibubbles}
Nicolas Moreno-Gomez, Athanasios~G Athanassiadis, Albert~T Poortinga, and Peer Fischer.
\newblock Antibubbles enable tunable payload release with low-intensity ultrasound.
\newblock {\em Advanced Materials}, 35(48):2305296, 2023.

\bibitem{kotopoulis2022formulation}
Spiros Kotopoulis, Christina Lam, Ragnhild Haugse, Sofie Snipstad, Elisa Murvold, T{\ae}raneh Jouleh, Sigrid Berg, Rune Hansen, Mihaela Popa, Emmet Mc~Cormack, et~al.
\newblock Formulation and characterisation of drug-loaded antibubbles for image-guided and ultrasound-triggered drug delivery.
\newblock {\em Ultrasonics Sonochemistry}, 85:105986, 2022.

\bibitem{EvonikAerosil}
Evonik.
\newblock {\em AEROSIL® fumed silica - Product overview}.
\newblock Evonik, Evonik Resource Efficiency GmbH Business Line Silica Rodenbacher Chaussee 4 63457 Hanau Germany, 2018.

\bibitem{melde:2016}
Kai Melde, Andrew~G. Mark, Tian Qiu, and Peer Fischer.
\newblock Holograms for acoustics.
\newblock {\em Nature}, 537(7621):518--522, September 2016.

\bibitem{OpenFOAM-Foundation2021}
{OpenFOAM-v2006}.
\newblock {https://www.openfoam.com/download/ release-history}, 2020.

\bibitem{stride2019nucleation}
Eleanor Stride and Constantin Coussios.
\newblock Nucleation, mapping and control of cavitation for drug delivery.
\newblock {\em Nature Reviews Physics}, 1(8):495--509, 2019.

\bibitem{shakya2024ultrasound}
Gazendra Shakya, Marco Cattaneo, Giulia Guerriero, Anunay Prasanna, Samuele Fiorini, and Outi Supponen.
\newblock Ultrasound-responsive microbubbles and nanodroplets: A pathway to targeted drug delivery.
\newblock {\em Advanced Drug Delivery Reviews}, page 115178, 2024.

\bibitem{wang2013ultrasound}
Tzu-Yin Wang, Katheryne E~Wilson, Steven Machtaler, and Jurgen K~Willmann.
\newblock Ultrasound and microbubble guided drug delivery: mechanistic understanding and clinical implications.
\newblock {\em Current pharmaceutical biotechnology}, 14(8):743--752, 2013.

\bibitem{delaney2022making}
Lauren~J Delaney, Selin Isguven, John~R Eisenbrey, Noreen~J Hickok, and Flemming Forsberg.
\newblock Making waves: how ultrasound-targeted drug delivery is changing pharmaceutical approaches.
\newblock {\em Materials Advances}, 3(7):3023--3040, 2022.

\bibitem{melde2024ultrasound}
Kai Melde, Athanasios~G Athanassiadis, Dimitris Missirlis, Minghui Shi, Senne Seneca, and Peer Fischer.
\newblock Ultrasound-assisted tissue engineering.
\newblock {\em Nature Reviews Bioengineering}, 2(6):486--500, 2024.

\end{thebibliography}
\end{footnotesize}
\newpage

\end{document}